\documentclass[showpacs,aps,graphicx,twocolumn]{revtex4}
\usepackage{graphicx}

\begin{document}

\title{Deterministic entanglement distillation for secure double-server blind quantum computation}

\author{Lan Zhou,$^{1,2}$ Yu-Bo Sheng,$^{2,3}$\footnote{shengyb@njupt.edu.cn} }
\address{ $^1$College of Mathematics \& Physics, Nanjing University of Posts and Telecommunications, Nanjing,
210003, China\\
$^2$Key Lab of Broadband Wireless Communication and Sensor Network
 Technology,
 Nanjing University of Posts and Telecommunications, Ministry of
 Education, Nanjing, 210003,
 China\\
$^3$Institute of Signal Processing  Transmission, Nanjing
University of Posts and Telecommunications, Nanjing, 210003,  China\\}

\date{\today }
\begin{abstract}
Blind quantum computation (BQC) provides an efficient method for the client who does not have enough
sophisticated technology and knowledge to perform universal quantum computation. The single-server BQC protocol requires the
client to have some minimum quantum ability, while
the double-server BQC protocol makes the client's device completely classical, resorting to the pure and clean Bell-state shared by  two servers.
In this paper, we provide a deterministic entanglement distillation protocol in a practical noisy environment for the double-server BQC protocol.
This protocol can obtain the pure maximally entangled Bell state with the success probability of 100\% in principle.
The distilled maximally entangled states can be remaind to perform the BQC protocol subsequently. The parties who perform the distillation protocol do not need to exchange
the classical information and they learn nothing from the client.  It makes this protocol unconditionally secure  and suitable for current BQC protocol.
\end{abstract}
 \maketitle

\section{Introduction}
Blind quantum computation (BQC) is a new type of quantum computation model which
can release the client who does not have enough knowledge and sophisticated technology
to perform the  universal quantum computation \cite{blind1,blind2,blind3,blind4,blind5,blind6,blind7,blind8,blind9,blind10,blind11,bllind12}. A complete quantum computation comprises two
parts. One is the client, say Alice, who has a classical computer and some ability of
quantum operation, or she may be completely classical. The other is the fully-fledged quantum computer server owned by Bob.
The first BQC protocol was proposed by Childs  in 2005 \cite{blind1}. It requires the standard quantum circuit model. In his protocol,
Bob needs to perform the quantum gates and Alice requires the quantum memory. In 2006, Arrighi and Salvail proposed another
BQC protocol where Alice needs to prepare and measure multiqubit entangled states. It is cheat sensitive for Bob  obtaining
some information, if he does not mind being caught \cite{blind2}. In 2009, Broadbent, Fitzsimons, and Kashefi proposed a different BQC model (BFK protocol) based on the one-way quantum computation \cite{blind3,one-way}. In their protocol,
Alice only requires to generate the single-qubit quantum state and a classical computer. She does not need the quantum memory. Moreover,
Bob cannot learn anything from Alice's input, output and her algorithm, which makes it unconditionally secure. Inspired by the BFK protocol, several BQC protocols have been proposed. For instance, Morimae \emph{et al.} proposed two BQC protocols based on the Affleck-Kennedy-Lieb-Tasaki state \cite{blind4}. Fitzsimons and Kashefi
 constructed
a new verifiable BQC protocol based on a new class of resource states \cite{bllind12}. Recently, Morimae and Fujii proposed a BQC protocol in which
Alice only makes measurements \cite{blind8}. The experimental realization of the BFK protocol based on the optical system was also reported \cite{blind10}.

Actually, the aim of the BQC is to let the client who does not have enough sophisticated quantum technology and knowledge perform the quantum computation.
Therefore,  the Alice's device and operation is more classical, the protocol is more successful. In BFK protocol, if Bob only has one service, Alice still needs some
quantum technology. On the other hand, if two servers are provided which are owned by Bob1 and Bob2, respectively, Alice does not require any quantum
technology. She can complete the quantum computation task with a classical computation, resorting to the classical communication. This protocol is called double-server BQC protocol. In double-server BQC protocol,  Bob1 and Bob2 should obey a
strong assumption that they cannot communicate  with each other. If not, they can learn the computation information from Alice and make the computation insecure. Before starting the BQC protocol, they should  share the maximally entangled Bell states. Unfortunately, in a realistic environment, the noisy channel  will greatly degrade the quality of the entanglement and it will make the whole protocol become a failure. Therefore, they should recover the mixed entangled states into the maximally entangled states.

 Entanglement purification is the standard way for distilling the high quality entangled state from low quality entangled state, which has been widely discussed
 in current quantum communication \cite{Bennett1,Bennett2,Deutsch,pan1,simon,shengpra1,shengpra2,lixh,wang1,loock,deng1}. In 1996, Bennett \emph{et al.} proposed the entanglement purification protocol (EPP) based on the controlled-not gate \cite{Bennett1}.
  In 2001, Pan \emph{et al.} proposed a novel EPP
  with linear optics  \cite{pan1}. There are some EPPs based on the nonlinear optics and hyperentanglement \cite{simon,shengpra1,shengpra2,deng1,lixh}. Unfortunately, in a standard EPP, they all need the local operation and classical communication. As pointed out by Morimae and Fujii \cite{blind11}, it is not sevi-evident
that the security of the double-server BQC protocol is guaranteed, when use the entanglement distillation protocol into the double-server blind protocol.
\begin{figure}[!h]
\begin{center}
\includegraphics[width=8cm,angle=0]{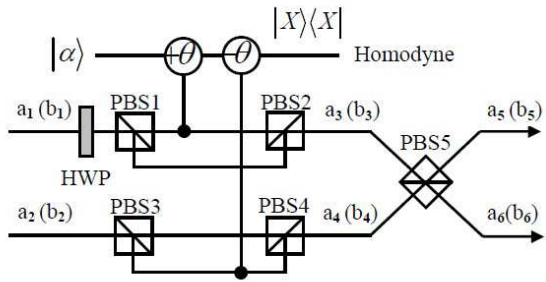}
\caption{Schematic of the principle of the quantum nondemolition (QND)
measurement  constructed by the cross-Kerr nonlinearity. HWP is the half wave plate which can make $|H\rangle\leftrightarrow |V\rangle$.  PBS is
the polarization beam splitter. It can transmit the $|H\rangle$ polarized photon and reflect the $|V\rangle$ polarized photon. $|\alpha\rangle$ is the coherent state.}
\end{center}
\end{figure}

Recently, Morimae and Fujii presented  a secure entanglement distillation protocol   based on the
one-way hashing distillation method \cite{blind11}. In their protocol, Alice first randomly chooses a $2n$-bit
string $s_{1}$ and sends it to two Bobs, respectively. Then each Bob  performs certain local unitary operation  determined by $s_{1}$.
By measuring a qubit of the single pair, Alice can obtain a bit information from the remained mixed state ensembles. Therefore, by repeating this protocol,
they can obtain $nS(\rho)$ bits of information about the mixed states ensembles. At the end of distillation, they can share about $n-nS(\rho)$ pairs.
\begin{figure}[!h]
\begin{center}
\includegraphics[width=8cm,angle=0]{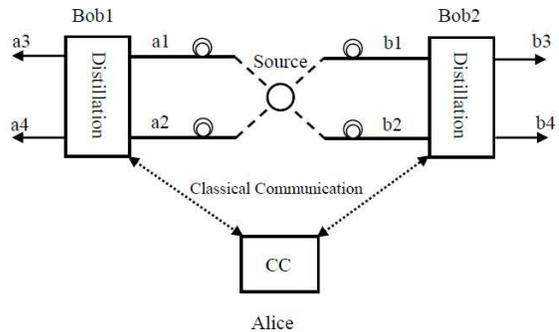}
\caption{Schematic of the principle of the double-sever BQC protocol combined with entanglement distillation. Bob1 and Bob2 can exchange the classical communication with Alice, respectively. But they cannot communicate with each other. CC is a classical computer. Both Bobs own the distillation equipment as shown in Fig. 1.}
\end{center}
\end{figure}

In this paper, we will present another deterministic entanglement distillation protocol for secure double-server BQC protocol. The whole protocol is based on the optical system, as the photons are well controlled and manipulated \cite{blind10}.  \textbf{This protocol is quite different from the one-way hashing distillation model and we resort to the hyperentanglement to complete the distillation \cite{hyper1,hyper2,hyper3}.} After performing the protocol, Alice can obtain the exact Bell state deterministically, with the success probability of 100\%, in principle,  according to the Bobs's measurement results, while she does not feedback any information to Bobs, which makes this distillation absolutely secure.

\section{Deterministic entanglement distillation with hyperentanglement}

Before we start to explain this protocol, let us introduce the distillation
equipment shown in Fig. 1. It is the quantum nondemolition (QND) measurement  with the cross-Kerr nonlinearity. As pointed out by Refs. \cite{QND1,QND2}, the Hamiltonian of the system is $H=\hbar\chi a^{\dagger}_{s}a_{s}a^{\dagger}_{p}a_{p}$. Here the $a^{\dagger}_{s}$,$a_{s}$($a^{\dagger}_{p}$, $a_{p}$) are the creation and destruction operators of the signal (probe) mode. From Fig. 1, if a single photon with vertical  polarization  ($|V\rangle$) in the spatial mode $a1$ passes through the equipment, the polarization of the photon will be flipped to horizonal polarization ($|H\rangle$) by half-wave plate (HWP) and transmit through the polarization beam splitter (PBS). The single photon combined with the coherent state $|H\rangle|\alpha\rangle$ will interact with the cross-Kerr nonlinearity and become $|H\rangle|\alpha e^{i\theta}\rangle$. It is shown that the single photon state $|H\rangle$ is unaffected but the coherent state picks up a phase shift directly proportional to the number of the photons. By measuring the phase of the coherent state, one can construct a QND measurement for the single photons.

The basic principle of the double-server BQC protocol  combined with  distillation model is shown in Fig. 2. The source (trust center) first generates a pair of hyperentangled state in both polarization and
spatial modes, which can be written as
\begin{eqnarray}
|\psi\rangle=\frac{1}{2}(|H\rangle|H\rangle+|V\rangle|V\rangle)\otimes(|a_{1}\rangle|b_{1}\rangle+|a_{2}\rangle|b_{2}\rangle).\label{hyper}
\end{eqnarray}
Such state is distributed to Bob1 and Bob2 through the spatial modes $a1$, $a2$, $b1$ and $b2$, respectively. \textbf{As pointed out by Refs.\cite{simon,shengpra2,deng1}, during the transmission,
the spatial entanglement  is more robust than polarization entanglement. Certainly, as pointed out by Simon and Pan \cite{simon}, the energy-time entanglement, which is more robust than the polarization entanglement and allows one to go to longer distances can also be used to perform this protocol.} The noisy channel will lead
the polarization part become a mixed state as
\begin{eqnarray}
\rho_{P}&=&F|\Phi^{+}\rangle\langle\Phi^{+}|+F_{1}|\Phi^{-}\rangle\langle\Phi^{-}|\nonumber\\
&+&F_{2}|\Psi^{+}\rangle\langle\Psi^{+}|+F_{3}|\Psi^{-}\rangle\langle\Psi^{-}|.
\end{eqnarray}
Here $F+F_{1}+F_{2}+F_{3}=1$. $|\Phi^{\pm}\rangle$ and $|\Psi^{\pm}\rangle$ are the polarized Bell states with
 \begin{eqnarray}
 |\Phi^{\pm}\rangle=\frac{1}{\sqrt{2}}(|H\rangle|H\rangle\pm|V\rangle|V\rangle),\nonumber\\
 |\Psi^{\pm}\rangle=\frac{1}{\sqrt{2}}(|H\rangle|V\rangle\pm|V\rangle|H\rangle).
\end{eqnarray}
 The whole system $\rho=\rho_{P}\otimes\rho_{S}$ can be described as a probabilistic mixture of four pure states:
with a probability of $F$, pair is in the state $|\Phi^{+}\rangle|\Phi^{+}\rangle_{s}$, with a probability $F_{1}$,
in the state  $|\Phi^{-}\rangle|\Phi^{+}\rangle_{s}$, with a probability of $F_{2}$, in the state $|\Psi^{+}\rangle|\Phi^{+}\rangle_{s}$,
and with a probability of $F_{3}$, in the state $|\Psi^{-}\rangle|\Phi^{+}\rangle_{s}$. Here $\rho_{S}=|\Phi^{+}\rangle_{s}\langle\Phi^{+}|$ with  $|\Phi^{+}\rangle_{s}=\frac{1}{\sqrt{2}}(|a_{1}\rangle|b_{1}\rangle+|a_{2}\rangle|b_{2}\rangle)$. After passing through the QNDs, the state $|\Phi^{+}\rangle|\Phi^{+}\rangle_{s}$ combined
with two coherent states $|\alpha\rangle_{B_{1}}$ and $|\alpha\rangle_{B_{2}}$ evolves as
 \begin{eqnarray}
&& |\Phi^{+}\rangle|\Phi^{+}\rangle_{s}|\alpha\rangle_{B_{1}}|\alpha\rangle_{B_{2}}\nonumber\\
 &=&\frac{1}{2}(|H\rangle|H\rangle+|V\rangle|V\rangle)
 \otimes(|a_{1}\rangle|b_{1}\rangle+|a_{2}\rangle|b_{2}\rangle)|\alpha\rangle_{B_{1}}|\alpha\rangle_{B_{2}}\nonumber\\
&\rightarrow&\frac{1}{2}(|H\rangle_{a_{1}}|H\rangle_{b_{1}}+|V\rangle_{a_{1}}|V\rangle_{b_{1}}\nonumber\\
&+&|H\rangle_{a_{2}}|H\rangle_{b_{2}}+|V\rangle_{a_{2}}|V\rangle_{b_{2}})|\alpha\rangle_{B_{1}}|\alpha\rangle_{B_{2}}\nonumber\\
&\rightarrow&\frac{1}{2}(|H\rangle_{a_{3}}|H\rangle_{b_{3}}|\alpha e^{i\theta}\rangle_{B_{1}}|\alpha e^{i\theta}\rangle_{B_{2}}\nonumber\\
&+&|V\rangle_{a_{3}}|V\rangle_{b_{3}}|\alpha\rangle_{B_{1}}|\alpha\rangle_{B_{2}}
+|H\rangle_{a_{4}}|H\rangle_{b_{4}}|\alpha\rangle_{B_{1}}|\alpha\rangle_{B_{2}}\nonumber\\
&+&|V\rangle_{a_{4}}|V\rangle_{b_{4}}|\alpha e^{-i\theta}\rangle_{B_{1}}|\alpha e^{-i\theta}\rangle_{B_{2}}).
\end{eqnarray}
The $|\alpha\rangle_{B_{1}}$ and $|\alpha\rangle_{B_{2}}$ are the coherent states used in the QND for Bob1 and Bob2, respectively. On the other hand, the  state $|\Psi^{+}\rangle|\Phi^{+}\rangle_{s}$ combined
with two coherent states $|\alpha\rangle_{B_{1}}$ and $|\alpha\rangle_{B_{2}}$ evolves as
 \begin{eqnarray}
&& |\Psi^{+}\rangle|\Phi^{+}\rangle_{s}|\alpha\rangle_{B_{1}}|\alpha\rangle_{B_{2}}\nonumber\\
 &=&\frac{1}{2}(|H\rangle|V\rangle+|V\rangle|H\rangle)
 \otimes(|a_{1}\rangle|b_{1}\rangle+|a_{2}\rangle|b_{2}\rangle)|\alpha\rangle_{B_{1}}|\alpha\rangle_{B_{2}}\nonumber\\
&\rightarrow&\frac{1}{2}(|H\rangle_{a_{1}}|V\rangle_{b_{1}}+|V\rangle_{a_{1}}|H\rangle_{b_{1}}\nonumber\\
&+&|H\rangle_{a_{2}}|V\rangle_{b_{2}}+|V\rangle_{a_{2}}|H\rangle_{b_{2}})|\alpha\rangle_{B_{1}}|\alpha\rangle_{B_{2}}\nonumber\\
&\rightarrow&\frac{1}{2}(|H\rangle_{a_{3}}|V\rangle_{b_{3}}|\alpha e^{i\theta}\rangle_{B_{1}}|\alpha\rangle_{B_{2}}\nonumber\\
&+&|V\rangle_{a_{3}}|H\rangle_{b_{3}}|\alpha\rangle_{B_{1}}|\alpha e^{i\theta}\rangle_{B_{2}}
+|H\rangle_{a_{4}}|V\rangle_{b_{4}}|\alpha\rangle_{B_{1}}|\alpha e^{-i\theta}\rangle_{B_{2}}\nonumber\\
&+&|V\rangle_{a_{4}}|H\rangle_{b_{4}}|\alpha e^{-i\theta}\rangle_{B_{1}}|\alpha\rangle_{B_{2}}).
\end{eqnarray}
If they consider the other items $ |\Phi^{-}\rangle|\Phi^{+}\rangle_{s}$ and $|\Psi^{-}\rangle|\Phi^{+}\rangle_{s}$, they can obtain the similar results.
Then Bob1 and Bob2 both measure the phase of the coherent state with the X quadrature measurement, which makes the $|\alpha e^{\pm i\theta}\rangle$  indistinguishable \cite{QND1}. Therefore, both Bobs only have two different results, say $\theta$ or 0. After the measurement, they both send their measurement results to Alice by classical communication. Finally, Alice can judge the exact Bell state according to the measurement results. In detail, if the measurement results are the same, say both $\theta$ or 0, they will obtain $|\Phi^{+}\rangle$, with the probability of $F+F_{1}$. Otherwise, if the measurement results are different, say Bob1 is $\theta$ and Bob2 is 0, or Bob1 is 0 and Bob2 is  $\theta$, they will obtain $|\Psi^{+}\rangle$, with the probability of $F_{2}+F_{3}$. During the whole protocol, two Bobs do not require to exchange their measurement results and they even do not know the information of the remained Bell state. They can only judge the output modes according to the different phase shift. If the coherent state picks up no phase shift, the photon must be in the upper output modes $a_{5}(b_{5})$. Otherwise, if the
coherent state picks up $\theta$ phase shift, the photon must be in the lower output modes  $a_{6}(b_{6})$.

Combined with the entanglement distillation, the double-server BQC protocol runs as follows:

Step 1: The entanglement source emits the hyperentangled pairs $|\psi\rangle$ to Bob1 and Bob2. They share $m$
pairs of mixed states $\rho^{\otimes m}$, because of the noise.

Step 2: Both Bobs perform the distillation protocol and send the measurement results to Alice. The purified states are $|\Phi^{+}\rangle^{\otimes[(F+F_{1})m]}$ and $|\Psi^{+}\rangle^{\otimes[(F_{2}+F_{3})m]}$.

Step 3: The following steps are the same as the traditional BQC protocol \cite{blind1,blind11}. Alice sends Bob1 classical messages $\{\theta_{j}\}^{m}_{j=1}$, where $\theta_{j}$ is randomly chosen by Alice from $\{\frac{k\pi}{4}|k=0,1,\cdots,7\}$. In detail, if Alice obtains $|\Phi^{+}\rangle$, she randomly sends Bob1 $\theta_{j}$, and if she obtains $|\Psi^{+}\rangle$, she randomly sends  $-\theta_{j}$.

Step 4: Bob measures his qubit in the $j$th Bell states in the basis $\{|0\rangle\pm e^{-j\theta_{j}}|1\rangle\}$ $(j=1,\cdots,m)$. Here we denote
$|H\rangle\equiv|0\rangle$ and $|V\rangle\equiv|1\rangle$. After Bob1 performing the measurement, he tells Alice the results $\{a_{j}\}^{m}_{j=1}$ with
$a_{j}\in\{0,1\}^{m}$.

Step 5: Alice and Bob2 start the single-server BQC protocol.

The traditional entanglement distillation protocols are unsuitable for double-server BQC protocol, because
message exchanges between two Bobs must be done through Alice's mediation. In this way, Bob1 can indirectly send a message to Bob2, which will make the computation insecure \cite{blind11}. Interestingly, this protocol does not require mediation. Alice can judge the deterministic Bell state according to the measurements results coming from two Bobs and start the BQC protocol subsequently.  During the total distillation, Alice does not feedback any messages to both Bobs. Once two Bobs learn nothing from Alice and cannot exchange the message with each other, it essentially means that distillation is absolutely secure.
 On the other hand,  both Bobs may have the evil intention and send  wrong messages to Alice. In this way, Alice will
obtain the wrong information about the Bell state, and it will induce the error computation. However, both Bobs still learn nothing from Alice.

\section{Discussion and conclusion}
Using spatial entanglement to purify the polarization entanglement has been studied for several groups \cite{simon,lixh,deng1}. However, their protocols
are all unsuitable for BQC protocol. In Ref. \cite{simon}, the bit-flip error can be well corrected by choosing the same output modes. However, they should require the traditional entanglement purification to correct the phase-flip error. In Refs. \cite{lixh,deng1}, with local operation and classical communication, both bit-flip error and phase-flip error can be corrected in one step. But the photon pair is destroyed because of the post-selection principle. In this protocol, the purified photon pair can be remained, resorting to the QND measurement. Moreover, both Bobs do not require to exchange the classical information, which makes it extremely suitable for double-server BQC protocol.
In a practical realization, they should generate the hyperentanglement and make the spatial entanglement  stable.  The generation of the hyperentanglement with both spatial and polarization degrees of freedom can be well solved with the spontaneous parametric down conversion (SPDC) source \cite{simon,pan1}. The pump
pulse of ultraviolet light passes through a $\beta$-barium borate
crystal (BBO). It can produce one pair of polarization entangled pairs with probability of $p$, and is
reflected and traverses the crystal a second time and can produce the same photon pairs  with the same order of magnitude. \textbf{This protocol realizes on the hypothesis that the spatial entanglement does not suffer from the noise. Though the spatial entanglement is robust than polarization entanglement, it still will be polluted in noisy channel. Interestingly, it usually suffers from the phase-noise, while the phase-noise can also be well controlled in current technology \cite{simon,pan1}. Moreover, the experiment for phase-noise measurements showed that the phase in long fibers (several
tens of km) remains stable, which is an acceptable level for
time on the order of 100 $\mu s$ \cite{phasenoise}.}
The other technology challenge may come from the cross-Kerr nonlinearity. Though many quantum information processes based on the cross-Kerr nonlinearity were proposed \cite{QND1,QND2,he1,lin1,shengpra1}, it is still a controversial topic \cite{Shapiro1,Gea}. Shapiro showed that single-photon Kerr nonlinearity may
do not help quantum computation \cite{Shapiro1}.  Gea-Banacloche also argued  that a large phase shift
via a "giant" Kerr effect with single-photon wave packets is
impossible\cite{Gea}. As pointed out by Kok \emph{et al.}, Kerr
phase shift is only $\tau\approx 10^{-18}$ in the optical single-photon
regime and a clean cross-Kerr nonlinearity is quite a
controversial assumption with current technology \cite{kok1}.  Fortunately, Hofmann showed that  a large phase shift of
$\pi$ can be obtained with a single
two-level atom in a one-sided cavity \cite{hofmann}. Using weak measurement, it is possible to amplify a cross-Kerr phase shift to an
observable value \cite{weak_meaurement}. The theoretical work of Zhu and Huang also showed that giant cross-Kerr nonlinearities
were also obtained in a double-quantum-well
structure with a four-level, double-type configuration \cite{oe}. The "giant" cross-Kerr effect with phase shift of 20 degrees per photon has been observed in current experiment \cite{gaint}.

In conclusion, we have presented a deterministic entanglement distillation protocol for double-server BQC protocol.  After performing the protocol, they can obtain the pure maximally entangled state with the success probability
of 100\% in principle. Bob1 and Bob2 do not communicate with each other and they also learn nothing from Alice. It makes the protocol unconditionally secure.

\section*{ACKNOWLEDGEMENTS}
This work is supported by the National Natural Science Foundation of
China under Grant No. 11104159, University Natural Science Research Project of Jiangsu Province
under Grant No. 13KJB140010, the open research fund of Key Lab of Broadband Wireless Communication and Sensor Network Technology, Nanjing University of Posts and Telecommunications, Ministry of Education (No. NYKL201303), Scientific
Research Foundation of Nanjing University of Posts and
Telecommunications under Grant No. NY213054, and a Project Funded by the Priority
Academic Program Development of Jiangsu Higher Education
Institutions.

\end{document}